\documentclass[10pt, conference, compsocconf]{IEEEtran}
\usepackage{graphicx}
\usepackage[lined,boxed,commentsnumbered, ruled]{algorithm2e}
\usepackage{amsmath}
\usepackage{multirow}
\usepackage{color}
\usepackage{url}
\usepackage{umlaute}
\usepackage{stfloats}
\usepackage{eurosym}

\ifCLASSINFOpdf
\else
\fi
\begin{document}
%
\title{Graph Clustering with Density-Cut}




%
\author{\IEEEauthorblockN{Junming Shao\IEEEauthorrefmark{1},
Qinli Yang\IEEEauthorrefmark{1},
Jinhu Liu\IEEEauthorrefmark{1} and
Stefan Kramer\IEEEauthorrefmark{2}}
\IEEEauthorblockA{\IEEEauthorrefmark{1}University of Electronic Science and Technology of China}
\IEEEauthorblockA{\IEEEauthorrefmark{2}University of Mainz}}


\maketitle

\begin{abstract}
How can we find a good graph clustering of a real-world network, that allows insight into its underlying structure and also potential functions? In this paper, we introduce a new graph clustering algorithm \emph{\textbf{Dcut}} from a density point of view. The basic idea is to envision the graph clustering as a \emph{density-cut} problem, such that the vertices in the same cluster are densely connected and the vertices between clusters are sparsely connected. To identify meaningful clusters (communities) in a graph, a density-connected tree is first constructed in a local fashion. Owing to the density-connected tree, \textbf{\emph{Dcut}} allows partitioning a graph into multiple densely tight-knit clusters directly. We demonstrate that our method has several attractive benefits: (a) \textbf{\emph{Dcut}} provides an intuitive criterion to evaluate the goodness of a graph clustering in a more natural and precise way; (b) Built upon the density-connected tree, \textbf{\emph{Dcut}} allows identifying the meaningful graph clusters of densely connected vertices efficiently; (c) The density-connected tree provides a connectivity map of vertices in a graph from a local density perspective. We systematically evaluate our new clustering approach on synthetic as well as real data to demonstrate its good performance.
\end{abstract}


%
\IEEEpeerreviewmaketitle

\section{Introduction}
\label{sec:intro}
Networks  now arise in various fields, e.g. social networks, protein-protein interaction networks, and the World Wide Web.  One of the key properties of these networks is community structure, interpreted as the presence of groups of nodes (called clusters or communities) with a high density of links between nodes in the same group, and a relatively low density of links between nodes in different groups \cite{Newman2002}. This compartmental organization of networks is ubiquitous in nature, such as group of friends in social networks, functional groupings in metabolic networks and different industrial sections in company networks. Exploring these clusters is crucial to understand the structural and functional properties of networks \cite{Lancichinetti2010}.

\begin{figure}[!t]
\centering
\includegraphics[height=30mm]{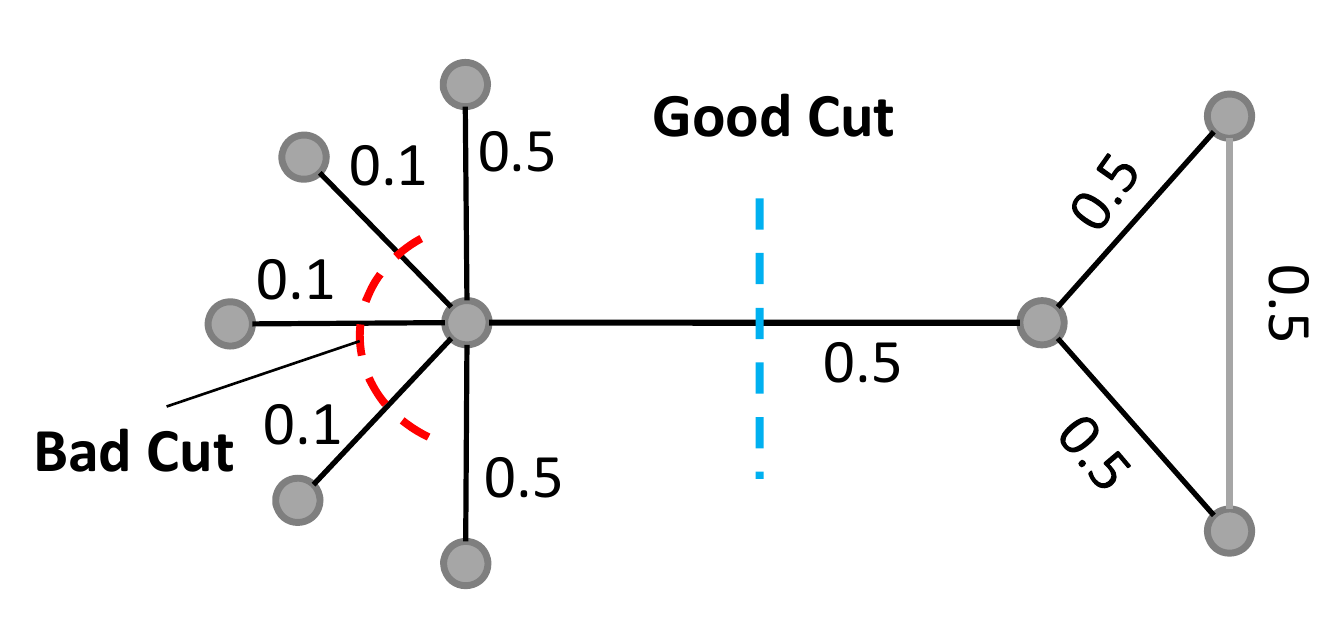}
\caption{A simple fictitious weighted graph. Here traditional graph clustering algorithms may produce a bad clustering, which is indicated by the red dashed line. A better partitioning should also consider the similarity of vertices in groups.}
\label{fig:Illustration}
\end{figure}

In recent years, the study of graph clustering has thus attracted a lot of attention, and many algorithms have been developed based on different criteria, e.g. \emph{betweenness} \cite{Newman2002}, \emph{normalized cut} ($Ncut$) \cite{ncut}, \emph{minimum-cut tree} \cite{Flake2004}, \emph{modularity} \cite{Newman2006}, to mention a few. Although many established approaches have already achieved some success, finding the intrinsic clusters in complex networks is still a big challenge \cite{Evans2010}. Up to now, most previous studies struggle to find a good graph clustering by minimizing the similarity between clusters, e.g. \emph{minimum cut} or \emph{modularity}. However, the similarities of vertices in a graph are considered only little.  For example, in Fig. \ref{fig:Illustration}, we show that the typical graph clustering algorithms, like \emph{Ncut}, may produce a bad grouping without considering the topological similarities among nodes. A good cut should minimize the similarities of vertices between clusters while maximize the similarities of vertices within each cluster.

Therefore, to find a good graph clustering, we try to answer the following two questions.

\vspace{2mm}
\textbf{ Q1: Quality} What is a natural and precise criterion to

\hspace{8mm}quantify the ``goodness" of a graph clustering?   \vspace{2mm}

\hspace{1mm}\textbf{Q2: Efficiency}  How can we manage to produce a good

\hspace{8mm}graph clustering efficiently? \vspace{2mm}

In this paper, we introduce a new density-based criterion for measuring the ``goodness" of a graph clustering. The basic idea is to consider the graph clustering as a density-cut problem by removing the edges in a proposed density-connected tree (cf. Section \ref{sec:tree}).  We expect that the vertices of resulting clusters are densely connected while the vertices between clusters are sparsely linked. The ``good cut" in Fig. \ref{fig:Illustration} is viewed as ``good", because the vertices in the same group have the same or similar topological structures, instead of focusing on minimum cuts (e.g. \emph{normalized cut}, \emph{ratio cut}) or expected cuts (e.g. \emph{modularity}) between two partitions only. Our proposal is to measure similarities of vertices in and between graph clusters by constructing a density-connected tree, where any two adjacent vertices with highest similarity (i.e. with strong edge weight and similar topological structure) are densely linked together.  Based on the properties of the density-connected tree, a good graph partitioning with \emph{density-cut} criterion is efficiently identified.

\subsection*{Contributions}
\label{sec:contributions}
In this paper, we present a new graph clustering method $Dcut$, which partitions a graph from the density point of view. The major benefits of $Dcut$ can be summarized as follows:
\begin{enumerate}
\item \emph{Density-cut criterion.} $Dcut$ provides a density-cut criterion for graph clustering. The new criterion is capable of measuring the quality of a graph clustering in a more natural and precise way. \vspace{2mm}

\item \emph{Good partitioning.} $Dcut$ allows producing a good graph partitioning, thanks to the intuitive density-cut criterion. By characterizing the density of any two adjacent vertices in local fashion, $Dcut$ can easily partition a graph into multiple clusters with densely connected vertices.\vspace{2mm}

\item \emph{Intuitive graph structure.}  $Dcut$ generates an intuitive and interpretable density-connected tree, which provides a connectivity map of vertices in a graph from a local density perspective. \vspace{2mm}

\item \emph{Efficiency.} Due to the properties of the density-connected tree, $Dcut$ is time efficient and easily implemented.

\end{enumerate}

The remainder of this paper is organized as follows: In the following section, we briefly survey related work. Section \ref{sec:main} presents our algorithm in detail. Section \ref{sec:experiment} contains an extensive experimental evaluation, before we conclude in Section \ref{sec:conclusion}.

\section{Related Work}
\label{sec:relWork}
During the past several decades, many approaches have been proposed for graph clustering, such as \cite{Karypis98afast}, \cite{ncut}, \cite{Newman2006} etc. Due to space limitations, we can only review the closest approaches from the literature. For detailed reviews of graph clustering, please refer to \cite{Schaeffer2007}.

\textbf{Minimum Cut.} The minimum-cut criterion based graph clustering refers to a class of well-known techniques which seek to partition a graph into disjoint subgraphs such that the number of cuts across the subgraphs is minimized.  Wu and Leahy  \cite{mincut} has proposed a clustering method based
on such minimum cut criterion, where the cut between two subgraphs is computed as the total weights of the edges that have been removed. $k-$disjoint subgraphs are obtained by recursively finding the minimum cuts that bisect the existing segments. To avoid an unnatural bias towards splitting small-sized subgraphs based on the minimum-cut criterion,  \emph{ratio cut} \cite{ratiocut} has been introduced, and it uses the second smallest eigenvalue of the similarity matrix to find the suitable cut. In the same spirit, Shi and Malik \cite{ncut} has proposed the \emph{normalized cut}, to compute the cut cost as a fraction of the total edge connections to all the nodes in a graph. To optimize this criterion, a generalized eigenvalue decomposition was used to speed up computation time. In many cases, this class of graph clustering algorithms relying on the eigenvector decomposition of a similarity matrix (e.g. \emph{ratio cut} and \emph{Ncut}) is also called spectral clustering.

\textbf{Modularity.} Recently, modularity has been developed to measure the division of a network into communities. Unlike minimum-cut related approaches which investigate the number of edges or the total number of edge weights between two subgroups, modularity identifies a good cut by measuring the expected edges between clusters.  Modularity-based graph clustering methods \cite{Newman2004}, \cite{Newman2006} partition a network into groups to ensure the number of edges between two groups is significantly less than the expected edges.

\textbf{Multi-Level Clustering.}
Metis is a class of multi-level partitioning techniques proposed by Karypis and Kumar \cite{multilevel}, \cite{Karypis98afast}. Graph clustering starts with constructing a sequence of successively smaller (coarser) graphs, and a bisection of the coarsest graph is applied. Subsequently, a finer graph is generated in the next level based on the previous bisections. At each level, an iterative refinement algorithm such as Kernighan-Lin (KL) or Fiduccia-Mattheyses (FM) is used to further improve the bisection. A more robust overall multilevel paradigm has been introduced by Karypis and Kumar \cite{Karypis98afast}, which
presents a powerful graph coarsening scheme. It uses simplified variants of KL and FM to speed up the refinement without compromising the overall quality.

\textbf{Markov Clustering.}
The Markov Cluster algorithm (MCL)  \cite{mcl} is a popular algorithm used in life sciences based on the simulation of (stochastic) flow in graphs. The basic idea is that dense regions in sparse graphs correspond to regions in which the number of random walks of length $k$ is relatively large. MCL basically identifies high-flowing regions representing the graph clusters by using an inflation parameter to separate regions of weak and strong flow.

\section{Graph Clustering based on Density-Cut}
\label{sec:main}
In this section, we present the $Dcut$ algorithm for graph clustering. In the following, we start with introducing the basic idea, and then a similarity measure is proposed to capture the similarity between two adjacent nodes. Based on the similarity measure, a density-connected tree representing the density of vertices is presented in Section \ref{sec:tree}. In Section \ref{sec:algo} we discuss the algorithm $Dcut$ in detail, and analyze its time complexity in Section \ref{sec:time}. \vspace{-2mm} \vspace{2mm}

\subsection{A Density-based Criterion for Graph Clustering}
\label{sec:idea}
As stated in Section \ref{sec:intro}, we consider the problem of graph clustering from an intuitive perspective: \emph{density}. We expect to find a good clustering if the vertices in each cluster are densely connected and the vertices between clusters are sparsely linked. In contrast to previous graph clustering algorithms, which treat ``density" as the total number of links or edge weights in or between clusters, we consider whether vertices in and between clusters are densely connected based on the similarities among adjacent nodes in a local fashion. If the similarity between two adjacent vertices in a graph is high, they are viewed as densely connected, and vice versa.  To identify tight-knit clusters, a density-connected tree is further proposed to look into the connection densities of vertices in a whole graph, where the two adjacent vertices with highest similarity (e.g. having strong edge weight and similar topological structure), are linked together in the tree. Built upon the density-connected tree, a good graph clustering based on the density-cut criterion is easily identified. In the following, we will first elaborate on how to measure the similarities between adjacent nodes.

\subsection{Node Similarity Measure}
\label{sec:nodesimilarity}
For the purpose of graph clustering, a similarity measure needs to be defined so that similar nodes can be assigned into the same group from the density point of view. Unlike most previous graph clustering, which use edge weight to represent the similarity of two connecting nodes, we characterize the similarity between any two adjacent nodes by combining their edge weight and their local topological structures. But before that, we start with some necessary definitions. \vspace{2mm}

\hspace{-3mm}\textsc{\textbf{Definition 1}} \hspace{1mm} (\textsc{Undirected Weighted Graph })
 Let $G=(V,E,W)$ be an undirected weighted graph, where $V$ is the set of nodes, $E$ is the set of edges and $W$ is the corresponding set of weights. $ e = \{u,v\} \in E$ indicates a connection between the nodes $u$ and $v$.  $w(u,v)$ represents the weight of edge $e$.  $\forall e = \{u,v\} \in E, w(u,v) = 1$, in case of unweighted graph. \vspace{2mm}

\hspace{-4mm}\textsc{\textbf{Definition 2}} \hspace{1mm} (\textsc{Neighbors of vertex $u$})
Given an undirected graph $G=(V,E,W)$, the neighborhood of a node $u \in V$ is the set $\Gamma(u)$ containing node $u$ and its adjacent nodes.
 \begin{equation}
\Gamma(u) =\{v \in V | \{u,v\} \in E\} \cup \{u\}
\label{eq:nb}
\end{equation}

In this study, we use the Jaccard coefficient \cite{jaccard} to quantify their local topological similarity. Generally, the more common neighbors two adjacent nodes have, the more similar they are. As the Jaccard coefficient normalizes the number of common neighbors by the sum of the size of the two neighborhoods, it captures the local connectivity density of any two adjacent nodes in a graph well. Formally, the Jaccard coefficient is defined as follow. \vspace{2mm}

\hspace{-4mm}\textsc{\textbf{Definition 3}} \hspace{1mm} (\textsc{Jaccard coefficient})
Given a graph $G=(V,E,W)$,  the Jaccard coefficient of any two adjacent nodes $u$ and $v$ is defined as:
\begin{equation}
\rho(u,v) = \frac{|\Gamma(u) \cap \Gamma(v) |}{|\Gamma(u) \cup \Gamma(v) |}
\label{eq:jaccard}
\end{equation}

By considering the topological structure and edge weight together, finally we define the similarity of any two adjacent nodes as follows.  \vspace{2mm}

\hspace{-4mm}\textsc{\textbf{Definition 4}} \hspace{1mm} (\textsc{Node Similarity})
Given any two adjacent nodes $u$ and $v$ in the graph $G$, the similarity of the two nodes $u$ and $v$ is defined as:
\begin{equation}
s(u,v)=  \rho(u,v) * w(u,v)
\end{equation}

The similarity measure captures the structural closeness and connection intensity of two adjacent nodes simultaneously.

\subsection{Density-connected Tree}
\label{sec:tree}
 Based on the similarity measure, we construct a density-connected tree (DCT) for clustering. The basic idea is to generate a density-connected chain which connects all vertices in a local density fashion. In this tree, all vertices of an original graph are directly linked, and each edge is associated with a weight representing the density connection (similarity) between two connected nodes. It is expected that similar vertices are densely connected in the tree while vertices in different communities are lightly connected. Formally, the steps of constructing the density-connected tree are as follows.

 Let $G = (V,E,W)$ be the original graph. $T$ is the density-connected tree under construction, and $T = null$ at the beginning. During the generation phase, two separate sets of vertices are maintained, where the first set is the vertices that have been inserted into the tree, and the second set is potential vertices for next insertion. For each step, the algorithm selects one node from the second set which maximizes the similarity to one existing node in the constructed tree. Specifically, in the initial phase, $T$ is an empty set and the status of all vertices are set as $unchecked$. The construction starts with randomly selecting any node in $V$ (e.g. $u$). We set $u.checked = true$, $u.connect = null$, $u.density = null$, and insert $u$ into $T$ at the first step.  In the second step, we search all unchecked adjacent vertices of nodes (neighbors) in $T$, and find one node (e.g. $v$) which has the highest similarity to one node already in $T$ (e.g. $u$) according to definition 4. Subsequently, we set $v.checked = true$, $v.conncet = u$, $v.density = s(u,v)$, and further insert $v$ into $T$ which directly links to the node $u$. The second step is repeated until all vertices have been inserted into the tree. Formally, the algorithm is described in Algorithm \ref{alg:dct}.

\begin{algorithm}
\caption{$T = DCT(G)$}
\label{alg:dct}
\SetKwData{Index}{Index}

\KwIn{$G = (V, E, W)$}
\KwOut{$T$}
\BlankLine
$T = null$; \\
Set $\forall v \in V$ as unchecked ($v.checked= false$);\\
Randomly selected one node $u \in V$; \\
Set $u.checked$ = true; \\
$u.connect = null$, and $u.density = null$; \\
$T.insert(u)$; \\ \vspace{2mm}

\While{$T.size < V.size$}{
    $maxv = -1$; $p = null$; $q =null$;\\
    \For{$i = 1$ \KwTo $T.size$}{
           $u$ = $T.get(i)$; \\
        \For{$j = 1$ \KwTo $\Gamma(u).size$}{
            $v$ = $\Gamma(u).get(j)$;\\
            \If{$v.checked$ == false}{
                \If{$s(u,v) > maxv$}{
                    $maxv = s(u,v)$;\\
                    $p = v$;\\
                    $q = u$;\\
                }
            }
        }
    }
    $p.checked$ = true; \\
    $p.connect = q$; $p.density = maxv$;\\
    $T.insert(p)$;\\
}
\end{algorithm}

\begin{figure*}[!ht]
\centering
\begin{tabular}{cccc}
\includegraphics[width=50mm]{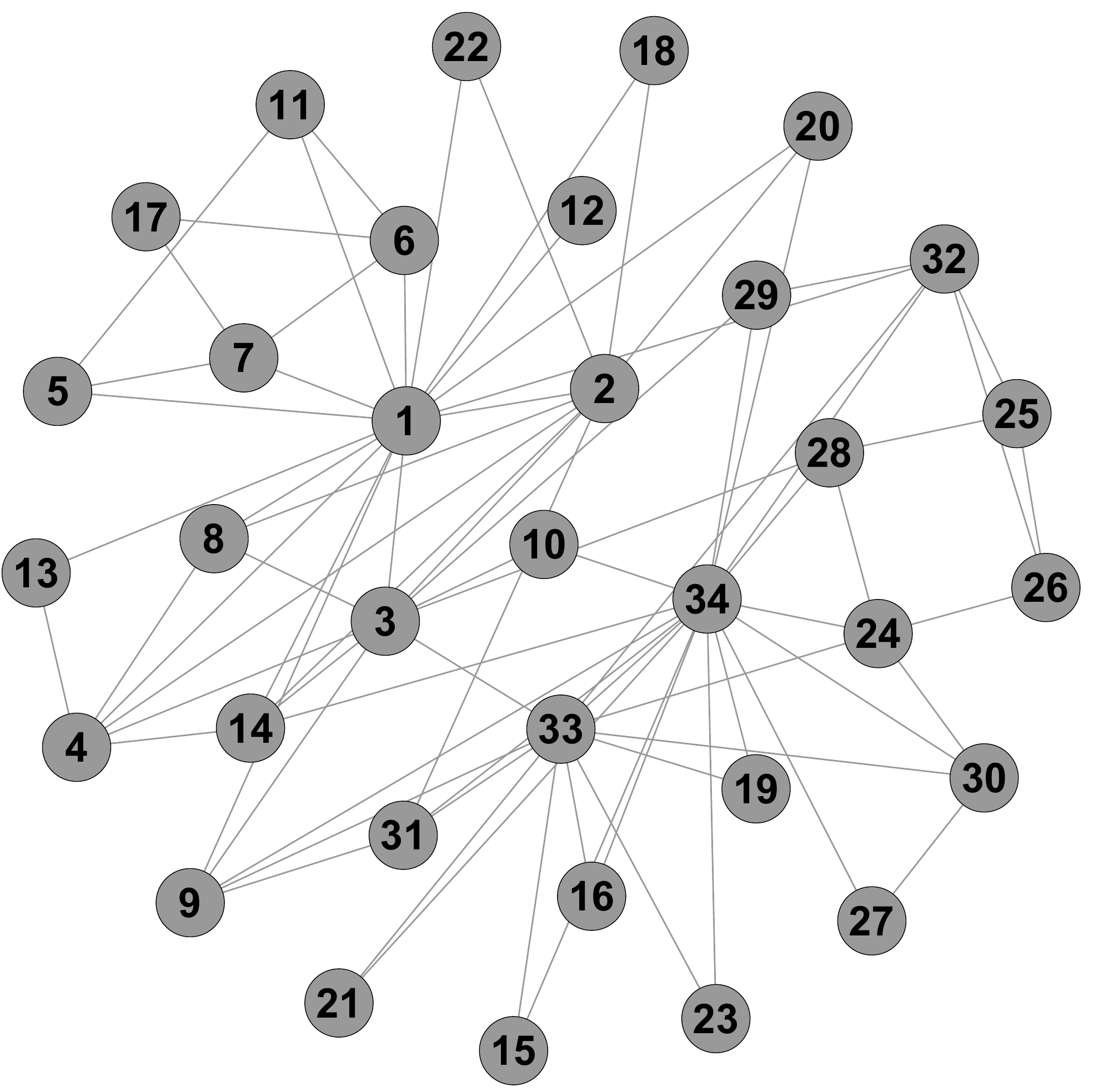}&
\includegraphics[width=84mm]{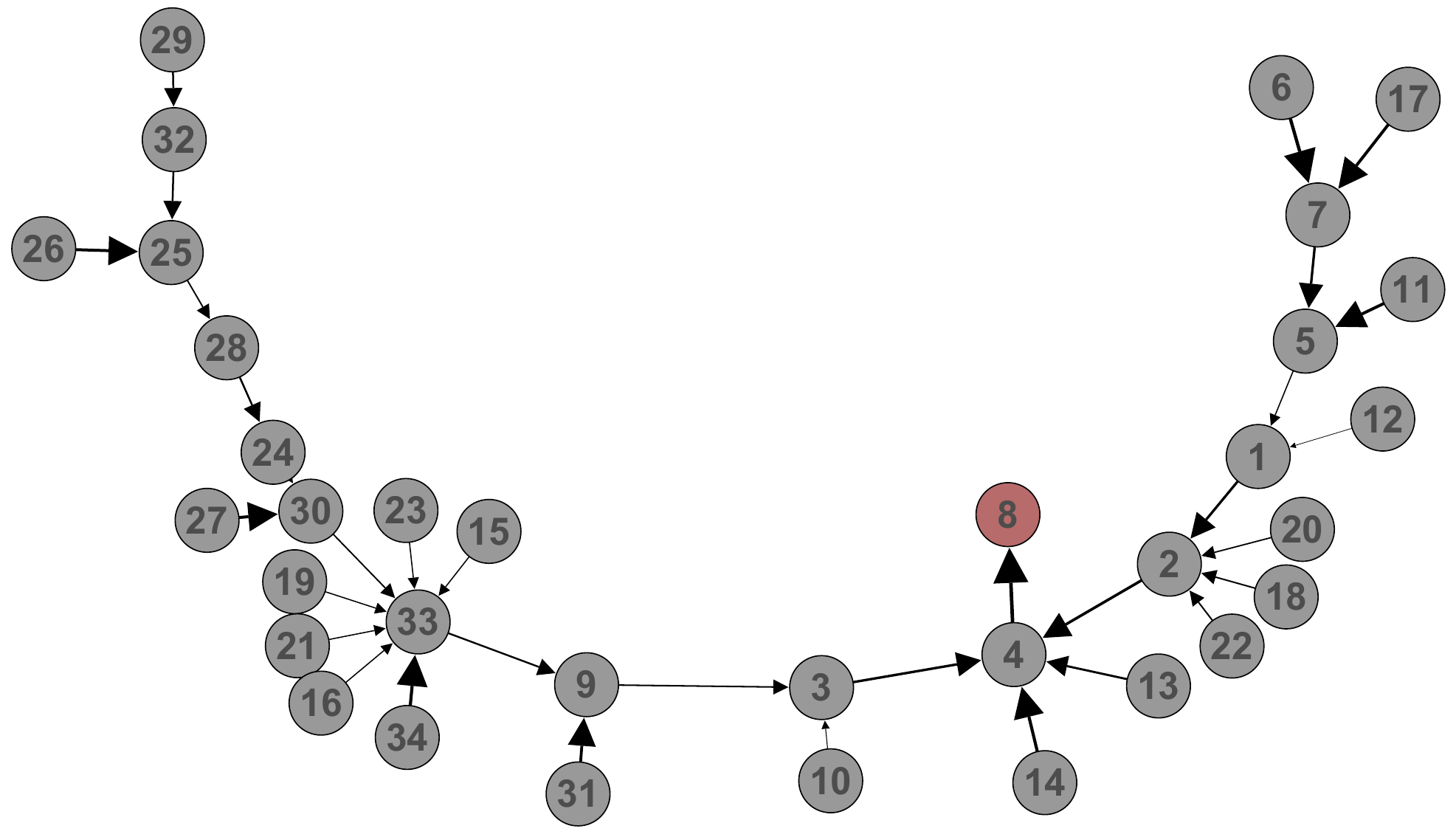}\\
(a) Zachary's Karate Club  & (b) Density-Connected Tree
\end{tabular}
\caption{The illustration of construction of the density-connected tree. The thickness of the arrows in the density-connected tree indicates how densely connected two nodes are.}
\label{fig:dct}
\end{figure*}

To illustrate DCT generation, Fig. \ref{fig:dct} takes the well-known Zachary's karate club network \cite{karate} as an example. The Karate graph data consists of 34 vertices and 78 undirected edges. Each node represents a member of the club, and each edge represents a tie between two members of the club. To construct its density-connected tree, one node is first randomly selected (e.g. node ``8" in this example, see Fig. \ref{fig:dct} (b)). Next, all unchecked adjacent vertices of node ``8'' are viewed as the potential vertices for next insertion, namely, the nodes of ``1", ``2", ``3" and ``4". As the node ``4" has the maximum similarity with node ``8", it is further inserted into the tree, which directly connects it to the node ``8" with the edge weight representing the similarity between the two nodes. For the next step,  as there are already two vertices (node ``8" and node ``4") in this tree,  all unchecked adjacent vertices are: $\Gamma(8) \bigcup \Gamma(4) \verb|\| \{4, 8\}$  (i.e. the nodes of  ``1", ``2", ``3", ``13" and ``14"). For the five potential vertices, the node ``14"  has the highest similarity with the node ``4" in the constructed tree, and thereby node ``14" is further inserted into the tree. Similarly, it directly connects to node ``4" with the corresponding similarity. This procedure is repeated until all vertices have been inserted into the tree.

In the density-connected tree, all vertices with the highest similarity are densely connected in a local fashion to form tight-knit components. It thus provides a summarization of the graph structure from the density perspective (Fig. \ref{fig:dct} (b)). Moreover, as the nodes with the highest similarity are linked together, the similarities of vertices in each component are maximized in this tree.\vspace{1mm}

\hspace{-4mm}\textbf{Theorem 1} \hspace{1mm} \emph{The density-connected tree (DCT) is unique for a given graph, if any two adjacent nodes have a distinct similarity.}

\hspace{-4mm}\textbf{Proof}. \hspace{1mm} Supposing there are two separate sets $R$ and $S$ during the generating phase at any step, where $R$ is the set of vertices that have been inserted into the tree, and $S$ is the set of unchecked neighbors of vertices that are already in $R$. The next node (e.g. $v$) is selected from $S$, which has maximum similarity with one node (e.g. $u$) from $R$. This means for each node, it is always connected with its most similar adjacent node. Since any two adjacent nodes in the graph have a distinct similarity, the connection of nodes $u$ and $v$ is unique. Thus, the density-connected tree for a given graph is unique.

Generally, DCT shows several desirable properties.

\begin{itemize}
\item \textbf{Sketch Graph}: DCT is a sketch graph that summarizes the original graph structure, and all vertices are connected without cycle. \vspace{2mm}
\item \textbf{Density Connectivity Map}:  DCT characterizes the density connectivity of vertices in graphs in a local fashion. It is intuitive that similar vertices are densely connected together, and vice versa. \vspace{2mm}

\item \textbf{Density Cut}: DCT offers an efficient way of identifying a good graph clustering, 
    which is further elaborated in the following Section \ref{sec:algo}. \vspace{2mm}
\end{itemize}


\subsection{The Dcut Algorithm}
\label{sec:algo}
In this section, we present $Dcut$ in detail. To find a good graph partitioning, as stated in Section \ref{sec:intro}, we consider graph clustering as a density-cut problem. Since DCT captures the density connectivity of vertices in a graph well, where the vertices with similar topological attributes and strong intensities of connections are densely linked together while the connections between components are lightly connected (the similarity of the two nodes connecting the two components is relatively low), it provides an intuitive solution to cut the edges in the tree directly to obtain a density-driven graph clustering.

Formally, we propose a new density-based criterion for measuring the ``goodness" of a graph clustering. Instead of investigating the value of total (or normalized) edge weights connecting the two partitions, our measure computes its density connection between the two partitions based on the density-connected tree. We call this measure the \emph{density cut} ($Dcut$):

\begin{equation}
Dcut(C_1,C_2) = \frac{d(C_1,C_2)}{min(|C_1|,|C_2|)},
\end{equation}

where $C_1, C_2$ are the two partitions, $d(C1,C2)$ means the corresponding density connecting the two partitions. The term of $min(|C_1|,|C_2|)$ is used to avoid the bias towards splitting small sets of vertices.

As DCT connects all vertices without cycle, each edge connects two components of a graph. Thereby, the intuitive bipartitioning of a graph in terms of density can be easily achieved by cutting one edge in the DCT. Instead of seeking to partition an original graph such that vertices in the same partition are densely connected and the vertices across different partitions are lightly connected, $Dcut$ allows recursively finding the optimal cut on the DCT directly.

Generally, supposing we want to partition a graph into $k$ disjoint clusters, the $Dcut$ algorithm runs in the following steps.

\begin{enumerate}
\item Given a graph G, compute the similarities between adjacent vertices based on the node similarity measure (Definition 4).

\item Construct its density-connected tree (see Alg. \ref{alg:dct}).

\item Partition the DCT by removing the edge with minimum \emph{Dcut} value of the resulting two components.

\item Recursively repartition the segmented DCT until $k$ components of the graph are obtained.
\end{enumerate}

\begin{figure}[!tb]
\centering
\begin{tabular}{cccc}
\includegraphics[width=80mm]{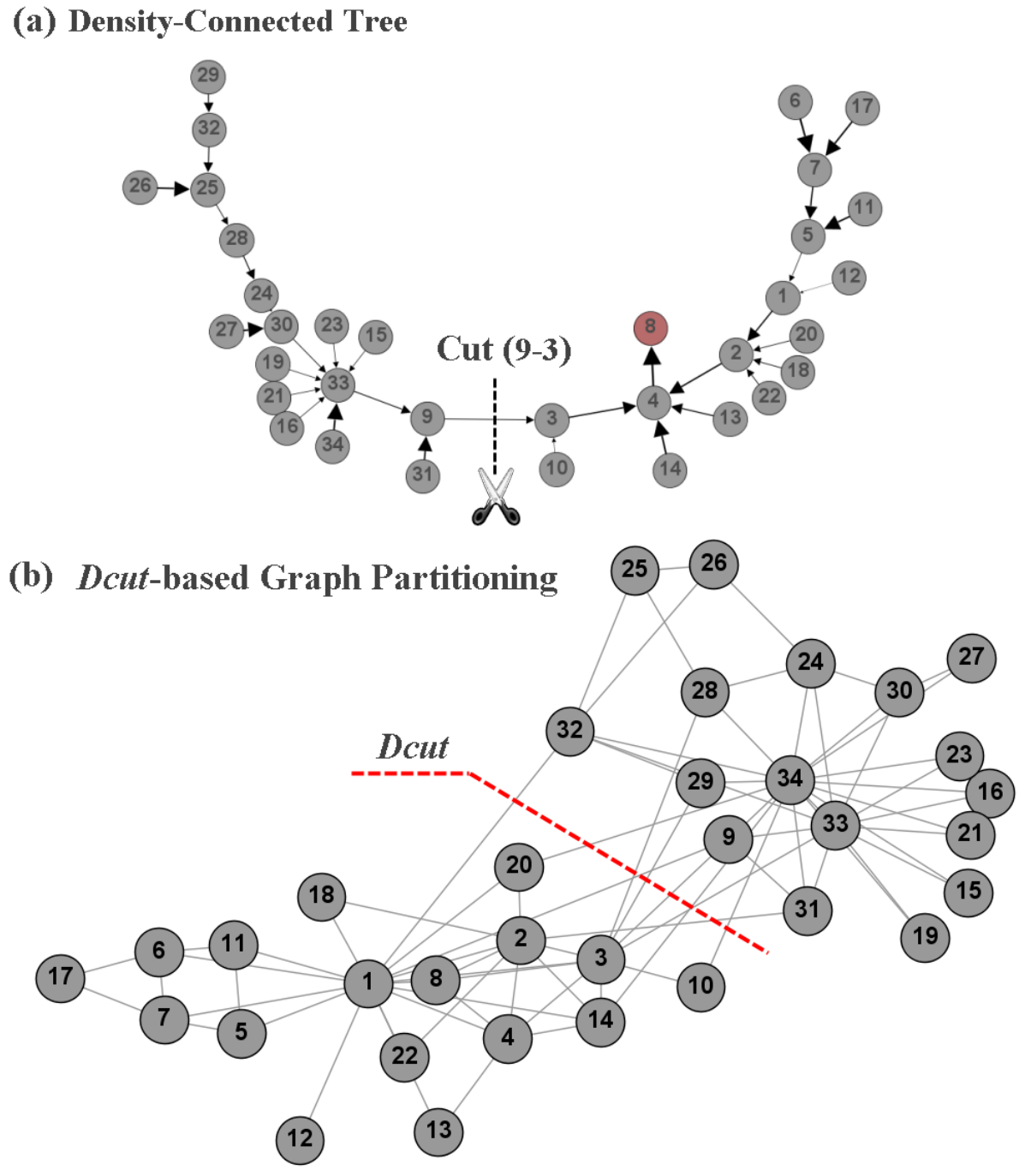}\\
\end{tabular}
\caption{The illustration of graph clustering based on the density-connected tree. }
\label{fig:dctcut}
\end{figure}

Fig. \ref{fig:dctcut} illustrates the graph clustering with density-cut criterion on the karate club network. Based on the constructed density-connected tree, the $Dcut$ values for cutting all edges are computed,  and the optimal cut with the minimum $Dcut$ value is found between node ``9" and node ``3". Removing this edge from the DCT results in the two partitions shown in Fig. \ref{fig:dctcut}.

%
%
%

\subsection{Complexity Analysis}
\label{sec:time}
To construct the density-connected tree, the similarities of all connecting vertices are first computed. The connecting vertices of each node have already been stored in the adjacency matrix. The computation of the similarity matrix in a graph requires $O(|E|)$ time. During the construction of the density-connected tree, for each step, all unchecked adjacent vertices are compared to find the maximum similarity. Therefore, the time complexity approximately
requires $\sum_{i=1}^{|V|}(i\cdot \frac{|E|}{|V|} - i) = (|E|-|V|)\cdot(1+|V|)/2$. Based on the density-connected tree, the recursive bipartitioning of the graph needs approximately $O(k\cdot|V|)$ times, where $k$ is the number of clusters. Finally, the time complexity of our algorithm is $O(|E|+k\cdot|V|+(|E|-|V|)\cdot(1+|V|))$.



\section{Experiments}
\label{sec:experiment}
In this section, we evaluate our proposed algorithm $Dcut$ on synthetic as well as real-world data to demonstrate its benefits.

\textbf{Comparison methods.} To examine the performance of $Dcut$, we compare it to two closely related graph clustering algorithms: the normalized cut criterion based graph clustering method $Ncut$ \cite{ncut} and the \emph{modularity}-based graph clustering algorithm by Newman \cite{Newman2006} (in the following named \emph{Modularity}). In addition, we compare to two representatives of graph clustering paradigms: the well-known multi-level partitioning algorithm $Metis$ by Karypis and Kumar \cite{Karypis98afast} and the Markov Cluster algorithm ($MCL$) by Dongen \cite{mcl}. In the experiments, $Dcut$, $Ncut$ and $Metis$ assume the same number of clusters $K$ for all data sets. $MCL$ takes the default inflation parameter as indicated in the original paper. All experiments have been performed on a workstation with 2.0 GHz CPU and 8.0 GB RAM.

\textbf{Evaluation measures:} To compare different graph clustering algorithms with respect to effectiveness, we evaluate the clustering results in two ways. First, if class label information is available for the graph, the clustering performance is directly measured by three widely used evaluation measures: \emph{Normalized Mutual Information (NMI)} \cite{nmi}, \emph{Adjusted Rand Index (ARI)} \cite{ari} and \emph{Cluster Purity}. All measures scale between 0 and 1 for a random or a perfect clustering result, respectively. For the graphs without ground truth, we adopt the well-known clustering coefficient proposed by Watts and Strogatz \cite{SW} as a cost function. This coefficient is a measure of the local cohesiveness that takes into account the importance of the clustered structure on the basis of the amount of triplets. Clustering results are measured by averaging the clustering coefficient of all subgraphs (clusters) obtained by different approaches. Formally, the clustering coefficient for measuring a graph clustering is defined as follows. \vspace{-1mm}
\begin{equation}
CC = \frac{1}{K}\sum_{i=1}^{K}\big{(}\frac{1}{|C_i|}\sum_{j=1}^{|C_i|} \frac{T_j}{L_j}\big{)},
\end{equation}

where $K$ is the number of clusters, $C_i$ is the $i-th$ cluster, $T_j$ is the number of triangles connected to node $j$, and $L_j$ is the number of triples centered around node $j$.

\begin{figure}[!tb]
\centering
\begin{tabular}{cccc}
\includegraphics[height=70mm]{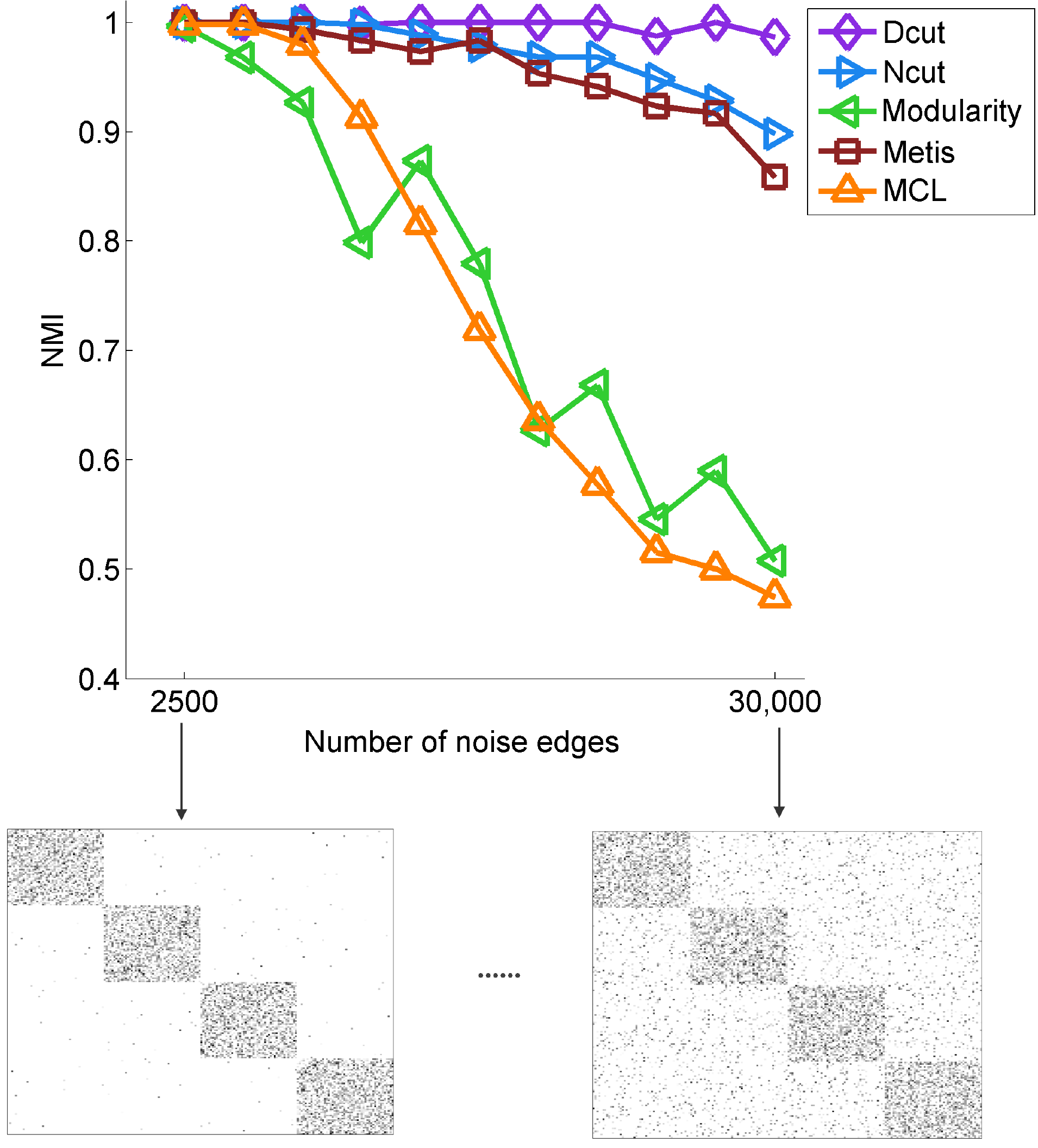}
\end{tabular}
\caption{Varying the number of inter-cluster edges in the data. Due to space limitations, the matrices only display 4 clusters, which is the same as in Fig. 5 and Fig. 6.}
\label{fig:noise}
\end{figure}

\begin{figure}[!tb]
\centering
\begin{tabular}{cccc}
\includegraphics[height=70mm]{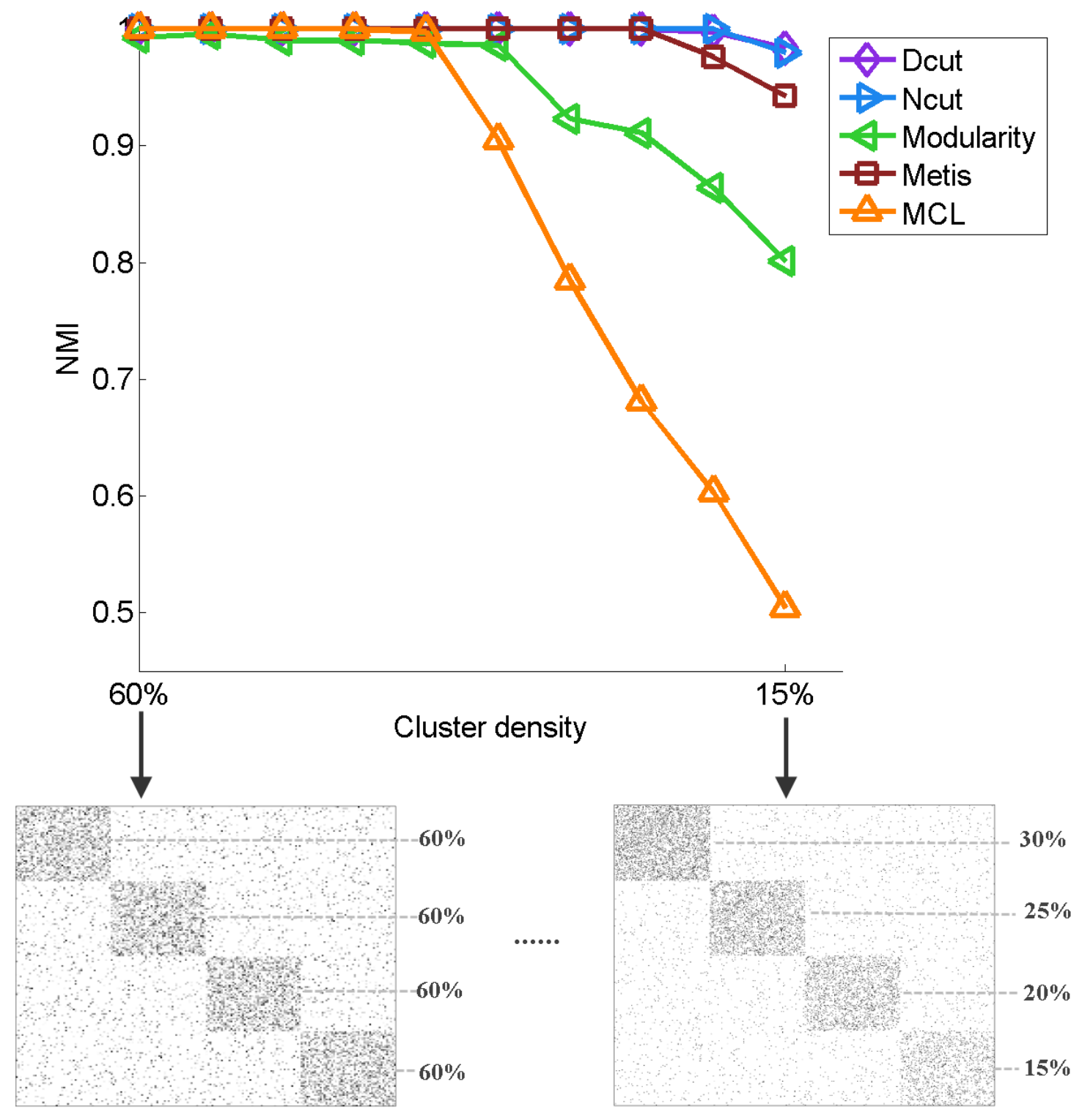}
\end{tabular}
\caption{Varying the densities of clusters in the data. }
\label{fig:density}
\end{figure}

\subsection{Synthetic data}
In this section, we start with several experiments on synthetic data sets featuring various graph characteristics.

\textbf{Noise Edges:} First, we evaluate how well the different graph clustering algorithms can handle the additional edges in graphs, which we call noise edges. Here 20 clusters are generated, and each cluster consisting of 50 nodes are randomly interlinked with 60\% intra-cluster edges. In addition to the approximately 15,000 intra-cluster edges, the number of noise edges, which are additional edges randomly added to random nodes, are present in the data varying from 2500 to 30,000. The noise in the data is represented by inter-cluster edges being added to the data, thus, introducing inter-cluster connectivity to hamper cluster separation.

 With adding more noise edges into the graph data (Fig. \ref{fig:noise}), the performance of all five approaches degrades, as measured by the normalized mutual information (NMI). $MCL$ is only able to handle data with up to approximately10,000 noise edges, and the performance starts to decrease dramatically as soon as more inter-edges are added. Like $MCL$, $Modularity$ is sensitive to noise edges, which is indicated by large performance fluctuations.  In contrast, the performances of $Dcut$, $Ncut$ and $Metis$ are more stable and robust to noise edges. $Dcut$ starts to achieve relatively better results than $Ncut$ and $Metis$ for up to 20,000 noise edges.
 \tabcolsep=0pt
\begin{figure*}[!tb]
\centering
\begin{tabular}{cccc}
\includegraphics[height=42mm]{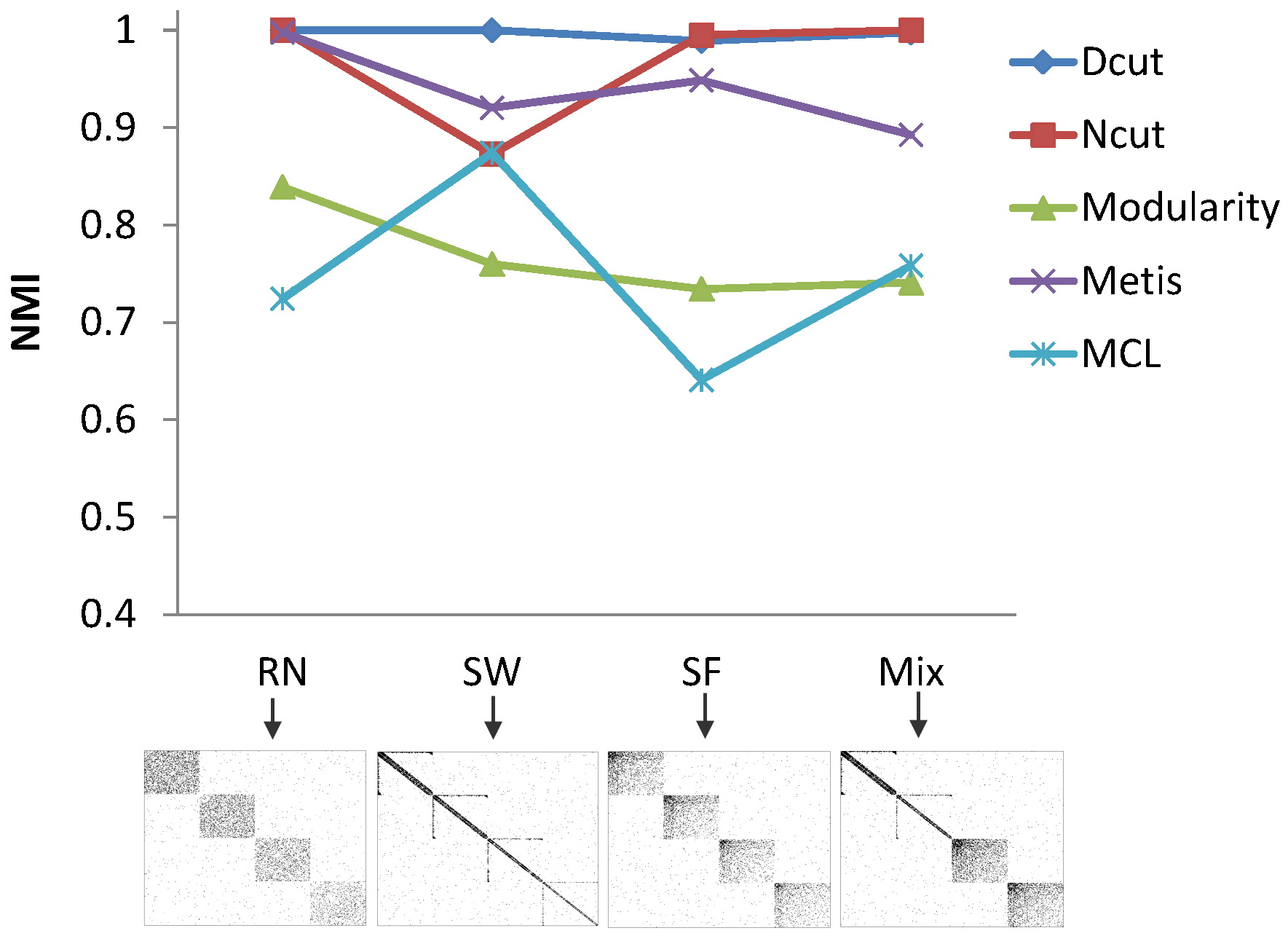}&
\includegraphics[height=42mm]{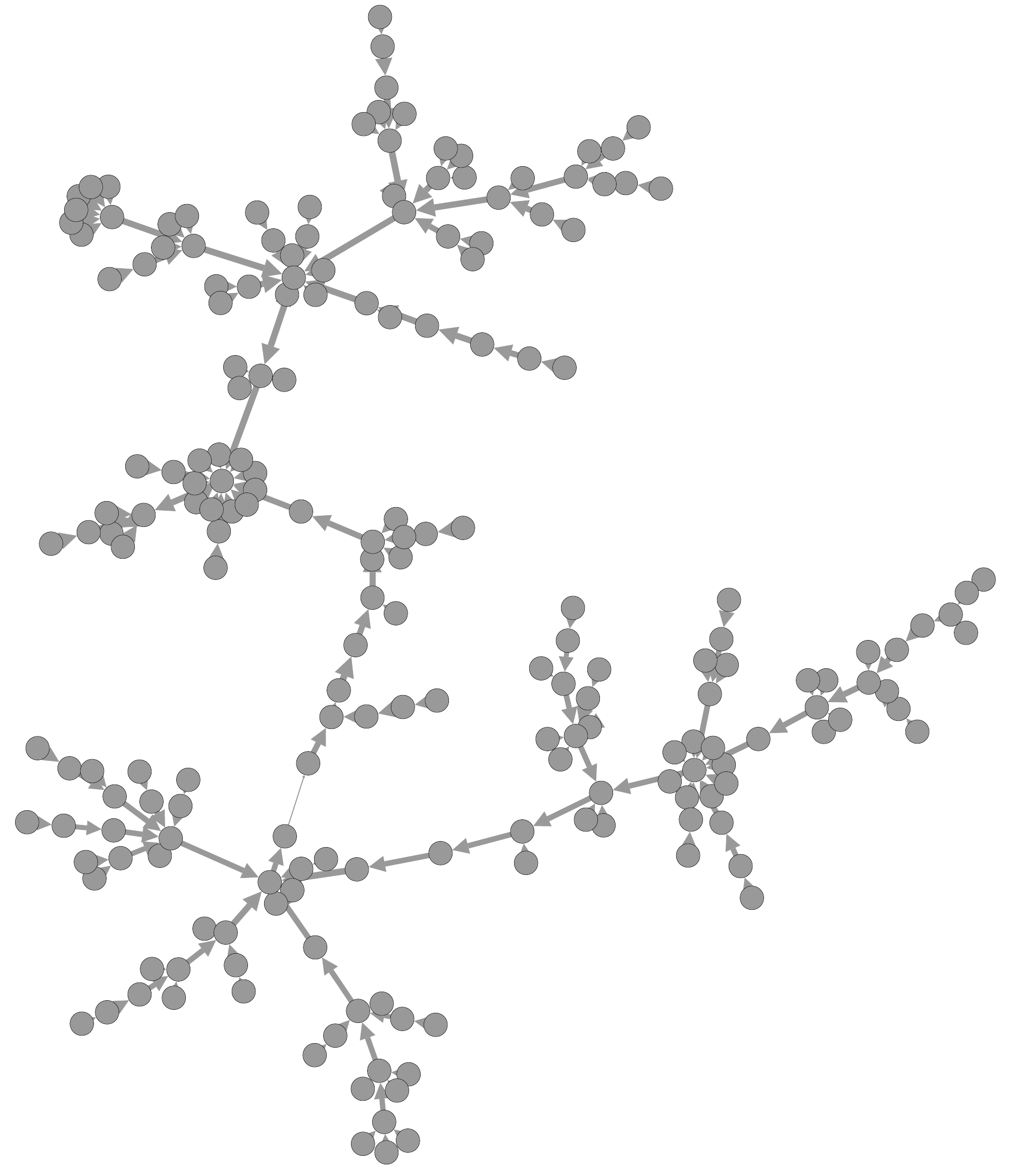}&
\includegraphics[height=38mm]{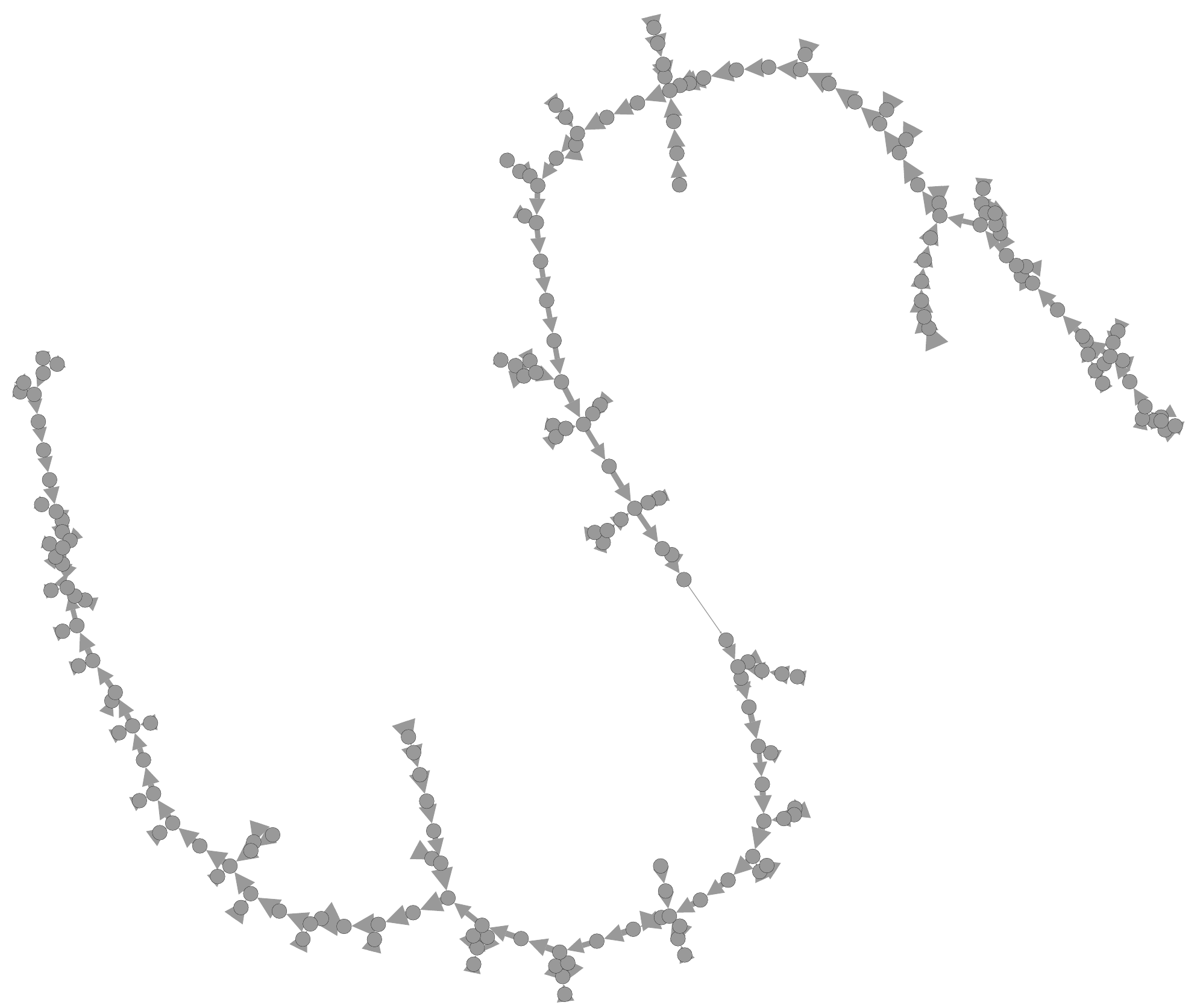}&
\includegraphics[height=40mm]{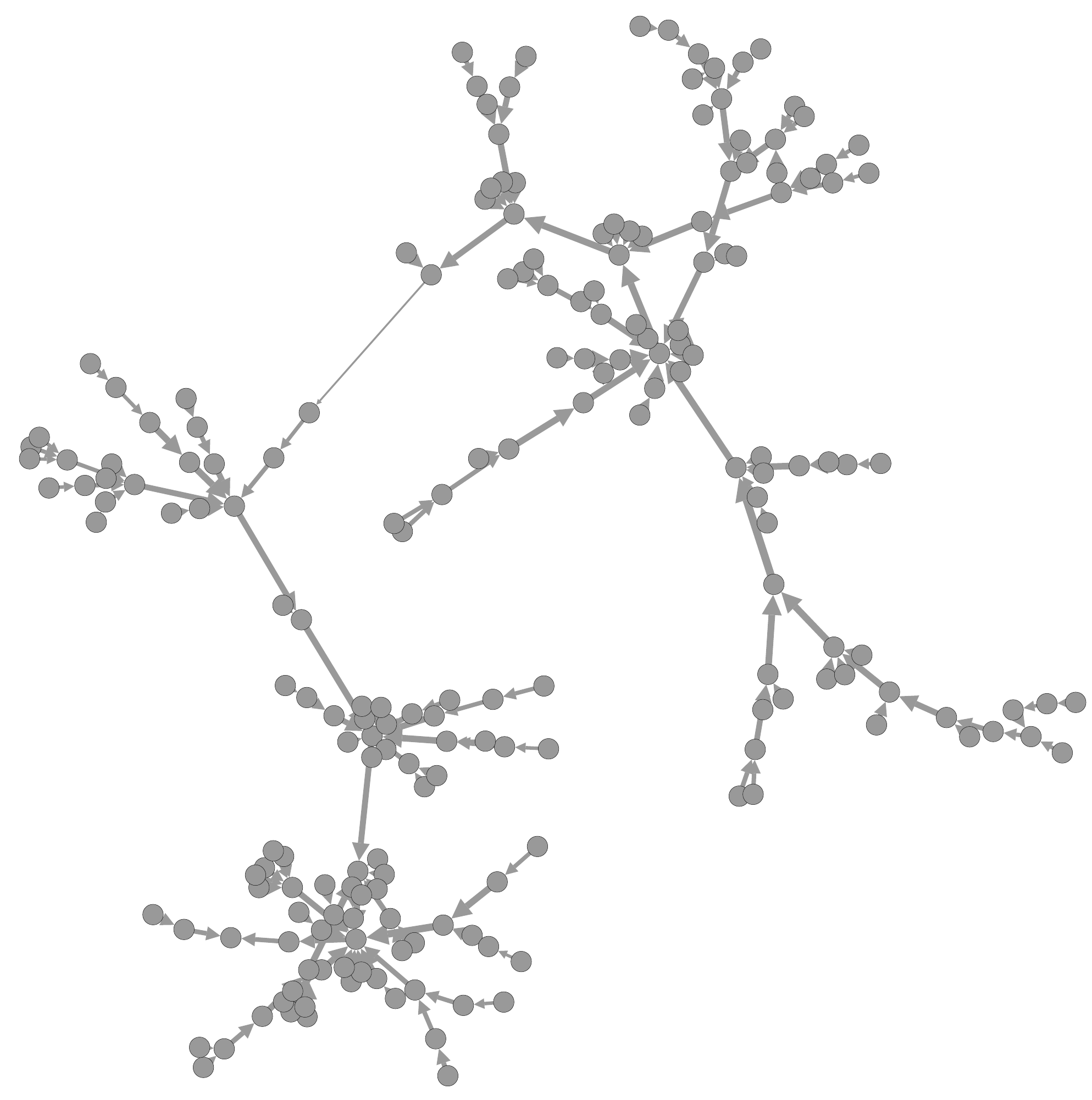}\\
 (a) Clustering results & (b) random network (RN)  &  (c) small-world network(SW) &  (d) scale-free network (SF)
\end{tabular}
\caption{Clustering on graph data of clusters characterizing different network types. Here, (b)- (c) are the density-connected trees corresponding different network types. Due to space limitations, only the graph data including two clusters is illustrated.}
\label{fig:graphtype}
\end{figure*}

 \textbf{Cluster Density:} Next, we evaluate how the algorithms respond to a change of the intra-cluster edges of different clusters in the graph data, which we call cluster density. Here 10 clusters are first generated with 5000 inter-cluster edges, and 100 nodes in each cluster are randomly interlinked with 60\% intra-cluster edges. We gradually change the number of intra-cluster edges for one cluster step by step with 5\% decrease until all clusters have different densities of intra-connectivity. As a result, the highest density of intra-cluster edges in the first cluster is 60\%, and the lowest density of intra-cluster edges in the last cluster is 15\% (Fig. \ref{fig:density}).

By generating the clusters with different densities in the graph data (Fig. \ref{fig:density}),  all algorithms perform well when the densities of clusters are above 50\%.  The performance of $MCL$ begins to decrease dramatically when clusters with lower densities are included in the graph data. $Modularity$ is also not able to achieve convincing results like $MCL$. As soon as the cluster density is lower than 35\%, $Metis$ starts to exhibit a slightly decreasing performance. Gradually changing the density of intra-cluster edges, both $Dcut$ and $Ncut$ achieve high clustering performance.

\begin{figure*}[!ht]
\centering
\begin{tabular}{cccc}
\includegraphics[height=60mm]{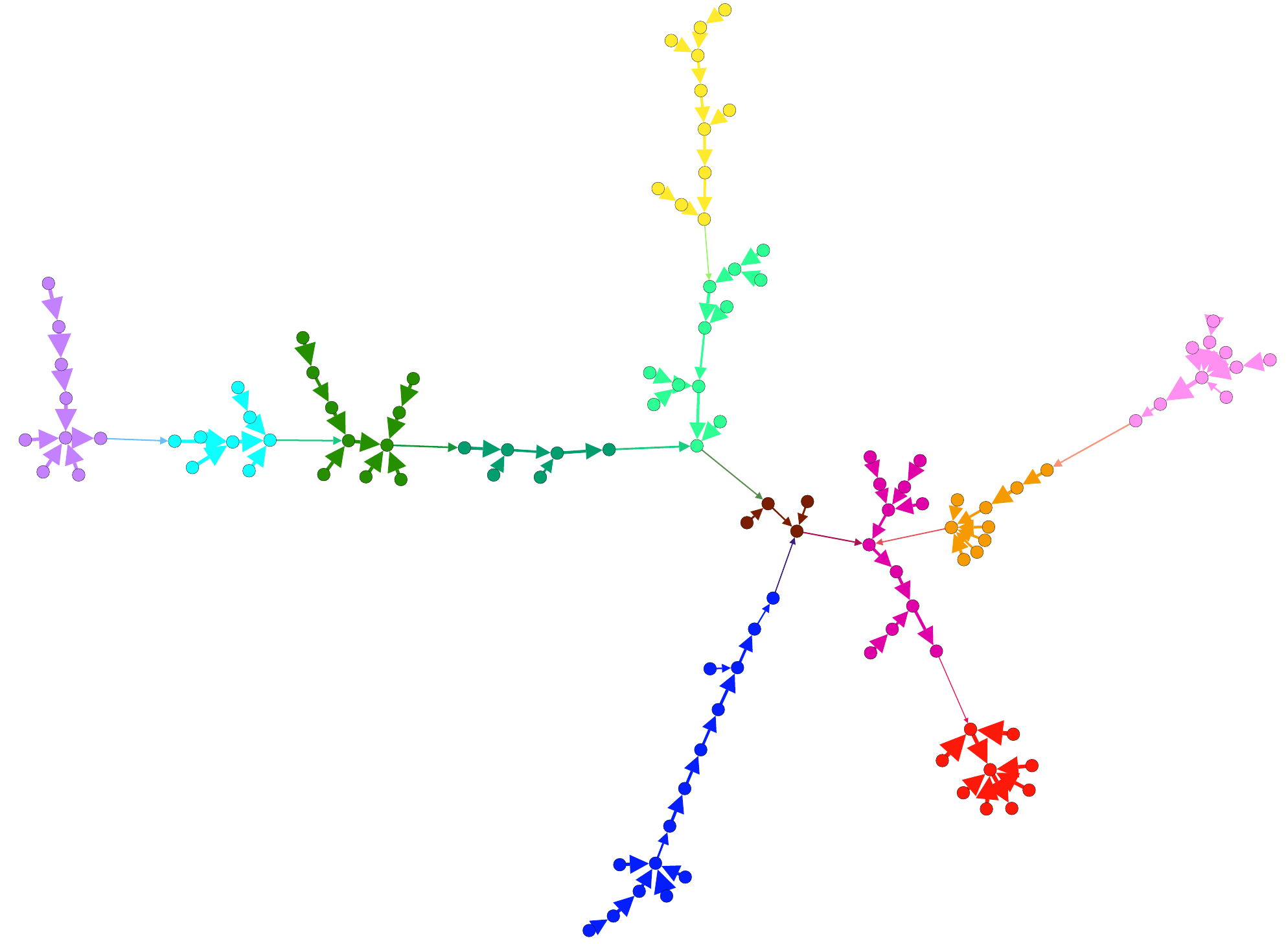}&
\includegraphics[height=64mm]{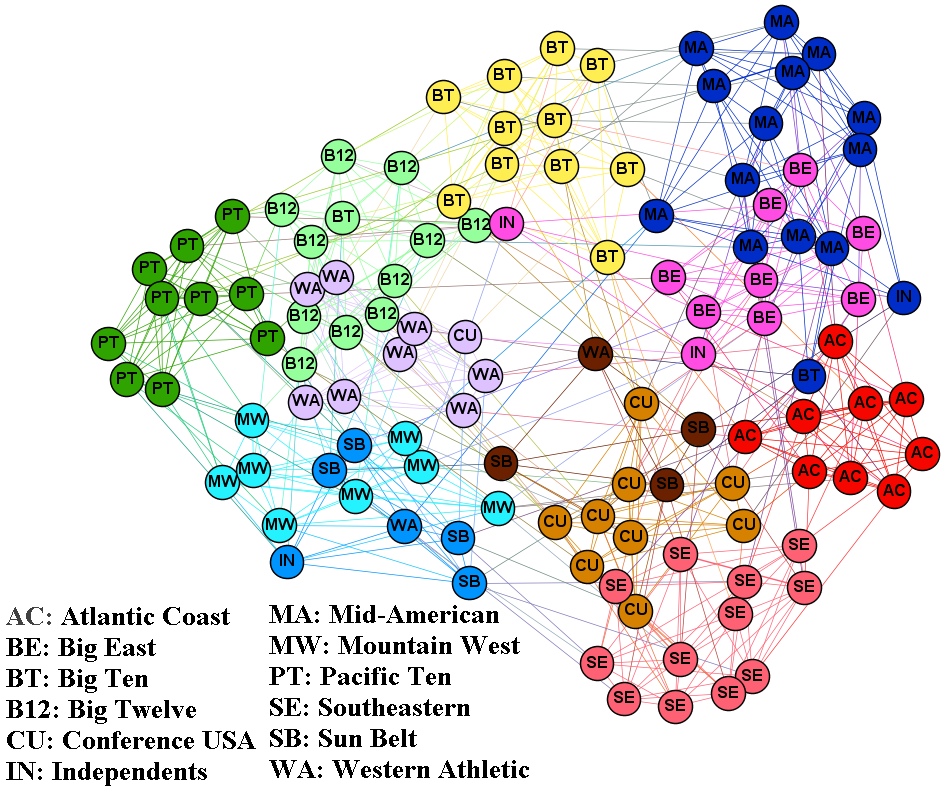}\\
(a) Density-connected Tree & (b) Graph partitioning with $Dcut$
\end{tabular}
\caption{Performance of $Dcut$ on the network of American college football, where the colors of nodes indicate different graph clusters.}
\label{fig:football}
\end{figure*}

 \begin{figure*}[!ht]
\centering
\begin{tabular}{cccc}
\includegraphics[height=40mm]{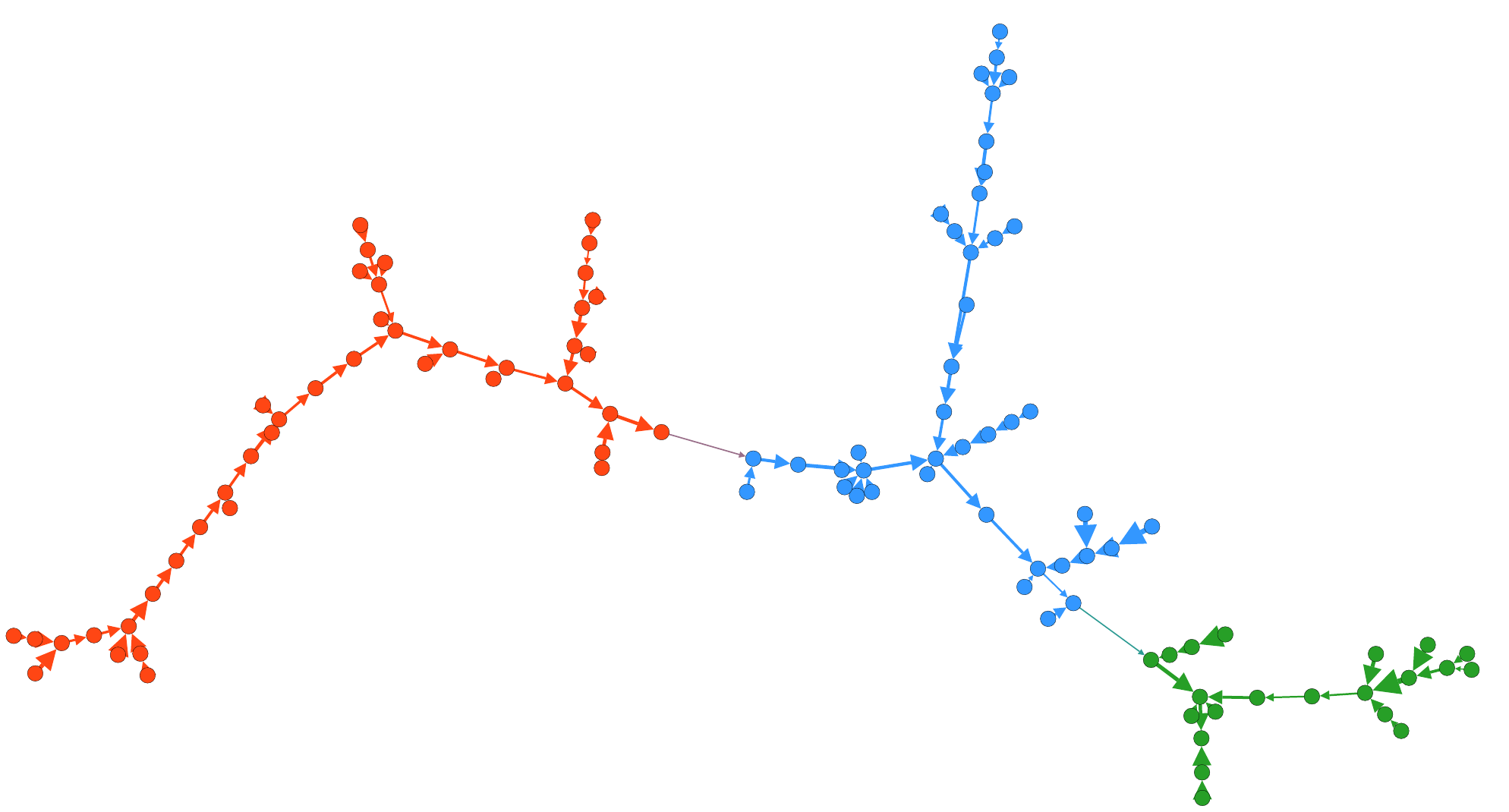}&
\includegraphics[height=40mm]{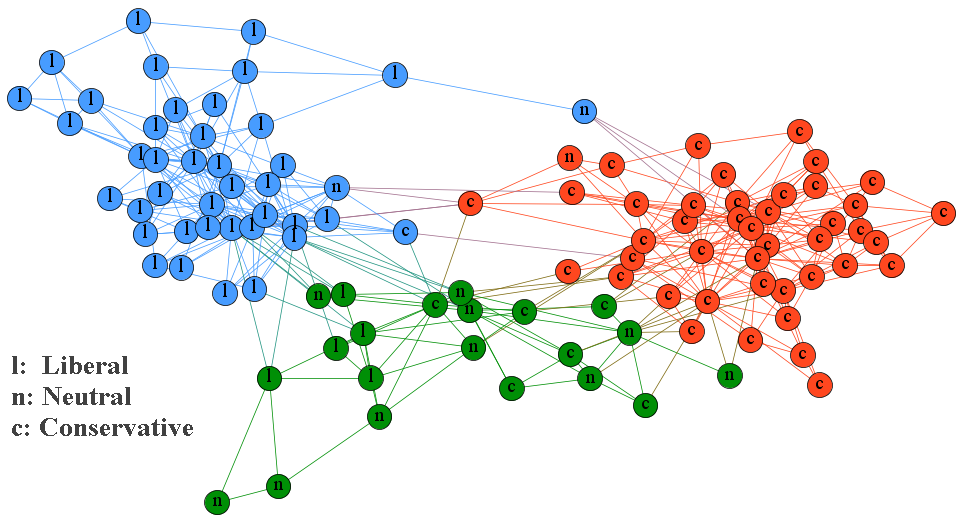}\\
(a) Density-connected Tree & (b) Graph partitioning with $Dcut$
\end{tabular}
\caption{Performance of $Dcut$ on the network of books about US politics.}
\label{fig:book}
\end{figure*}

\tabcolsep=4pt
\begin{table*}[!htb]
\center \caption{Performance of different graph clustering algorithms on real-world data sets.}
\begin{tabular}{|c|c|c|c|c|c|c|c|c|c|c|c|c|c|c|c|c|c|c|c|c|c|c|c|}
\hline \multirow{2}{*}{Data}  & \multicolumn{3}{|c|}{Dcut} & \multicolumn{3}{|c|}{Ncut} & \multicolumn{3}{|c|}{Modularity}& \multicolumn{3}{|c|}{Metis} & \multicolumn{3}{|c|}{MCL}\\ \cline{2-16}
   & NMI & ARI & Pur & NMI & ARI & Pur& NMI & ARI & Pur & NMI & ARI & Pur & NMI & ARI & Pur\\ \hline
College football &0.924  & 0.899 &0.930& 0.923 &0.897  &0.930 &0.596    &0.474&0.574&0.526&0.236&0.487&0.923&0.897&0.930\\ \hline
Politics Books  & 0.572 &0.680 &0.857 & 0.534 &  0.645 &0.829 &0.508&0.638&0.838&0.382&0.425&0.781&0.455&0.594&0.857\\ \hline
\end{tabular}
\label{tab:real1}
\vspace{-2mm}
\end{table*}

 \textbf{Network Types:} Finally, we evaluate how the algorithms depend on different types of networks. Here we generate several graph data of clusters representing different types of networks with various preferential attachments: random network, small-world network and scale-free network (Fig. \ref{fig:graphtype}). Specifically, the first graph data including 10 clusters representing a random network is generated, and each cluster consisting of 100 nodes is randomly interlinked with intra-cluster edges from the density of 55\% to 10\% with a stepwise decrease of 5\%. The second graph data consisting of 10 clusters exhibit small-world properties following the SW model \cite{SW}. Every node in each cluster connects to its $k$ nearest neighbours with rewiring probability of 20\%. $k$ ranges from 2 to 20 (step size of 2) for the different 10 clusters. Similarly, we generate the third graph data of 10 clusters representing a scale-free network \cite{BA} with various degrees of density. For each cluster of 100 nodes, the initial nodes are first generated and randomly linked, and for the next new node, $m$ edges are added to ensure that the degree distribution of each cluster follows the power law. The value of $m$ increases for clusters varying from 2 to 20. Moreover, graph data including clusters representing different types of networks are generated, where two clusters correspond to a random network, four clusters correspond to a small-world network and the remaining four clusters are scale-free networks. For all four graph data, 5000 inter-edges are additionally added.

 From Fig. \ref{fig:graphtype}(a), we can observe that most approaches perform well on the random network except for $MCL$. For the small-world network, the scale-free network and the mixed network, different methods exhibit different preferences. For all graph data, $MCL$ and $Modularity$ are not able to achieve convincing clustering results. $Dcut$ is the only algorithm which performs well for all graph data including mixed network types (Fig. \ref{fig:graphtype}(a)). To understand the reason behind that, we plot the corresponding DCTs for the graph data representing different network types.  For random network, vertices are connected with very similar density indicated by the similar thickness of arrows. The DCT of small-world networks are chain-like, while for scale-free networks, many vertices form a hub, and the strengths of connections decrease from the center to the outside. These features of DCTs capture the intrinsic characteristics of different network types well.

\subsection{Real World Data}
In this section, we evaluate the performances of different graph clustering algorithms on several real-world data which are all publicly available from the UCI network data repository (https://networkdata.ics.uci.edu/index.php).

\vspace{2mm}
\hspace{-4mm}\textbf{1. Networks with class information}
\vspace{2mm}

To provide an objective evaluation of different graph clustering algorithms, we first investigate the networks for which the ground truth of community structure are already known. The external evaluation measures such as \emph{NMI}, \emph{ARI} and \emph{purity} are applied.

 \textbf{American College football:} The graph data derived from the American football games of the schedule of Division I during regular season Fall 2000, where 115 vertices in the graph represent teams, and edges represent regular-season games between the two teams they connect. The teams are divided into 12 conferences containing around 8-12 teams each, and thereby the real community structure is already known.

$Dcut$ ($K$ = 12) identifies the graph clusters with a high degree of success (Fig. \ref{fig:football} and Table \ref{tab:real1}). Most teams are correctly grouped with the other teams in their conference with the highest cluster quality compared to the other four approaches (NMI = 0.924, ARI = 0.899, Purity = 93.0\%). The good performance is due to the density-connected tree, where the most closely associated teams are densely connected together (Fig. \ref{fig:football}(a)). $Ncut$ and $MCL$ also perform well, and most teams are correctly grouped. For $Metis$ and $Modularity$, however, it is difficult to discover the community structure. The performance of the different algorithms is summarized in Table \ref{tab:real1}.

 \textbf{Books about US politics:} The network consists of 105 nodes and 441 edges, which are derived from the books about US politics published around the time of the 2004 presidential election and sold by the online bookseller ``Amazon.com". Edges represent frequent copurchasing of books by the same buyers. Each book is categorized as ``liberal", ``neutral", or ``conservative" by Mark Newman based on a reading of the descriptions and reviews of the books posted on Amazon.  With $K=3$, most books can be correctly grouped by $Dcut$ with $NMI = 0.572$ (Fig. \ref{fig:book}). Two major clusters correspond to liberal and conservative books with high cluster purity (only four books are misclustered in the two clusters), respectively. The same types of books are linked together in the density-connected tree (Fig. \ref{fig:book}(a)). Compared with other algorithms, $Dcut$ achieves the best clustering results, as indicated in Fig. \ref{fig:book}(b) and Table \ref{tab:real1}.

\begin{figure}[!tb]
\centering
\begin{tabular}{cccc}
\includegraphics[height=56mm]{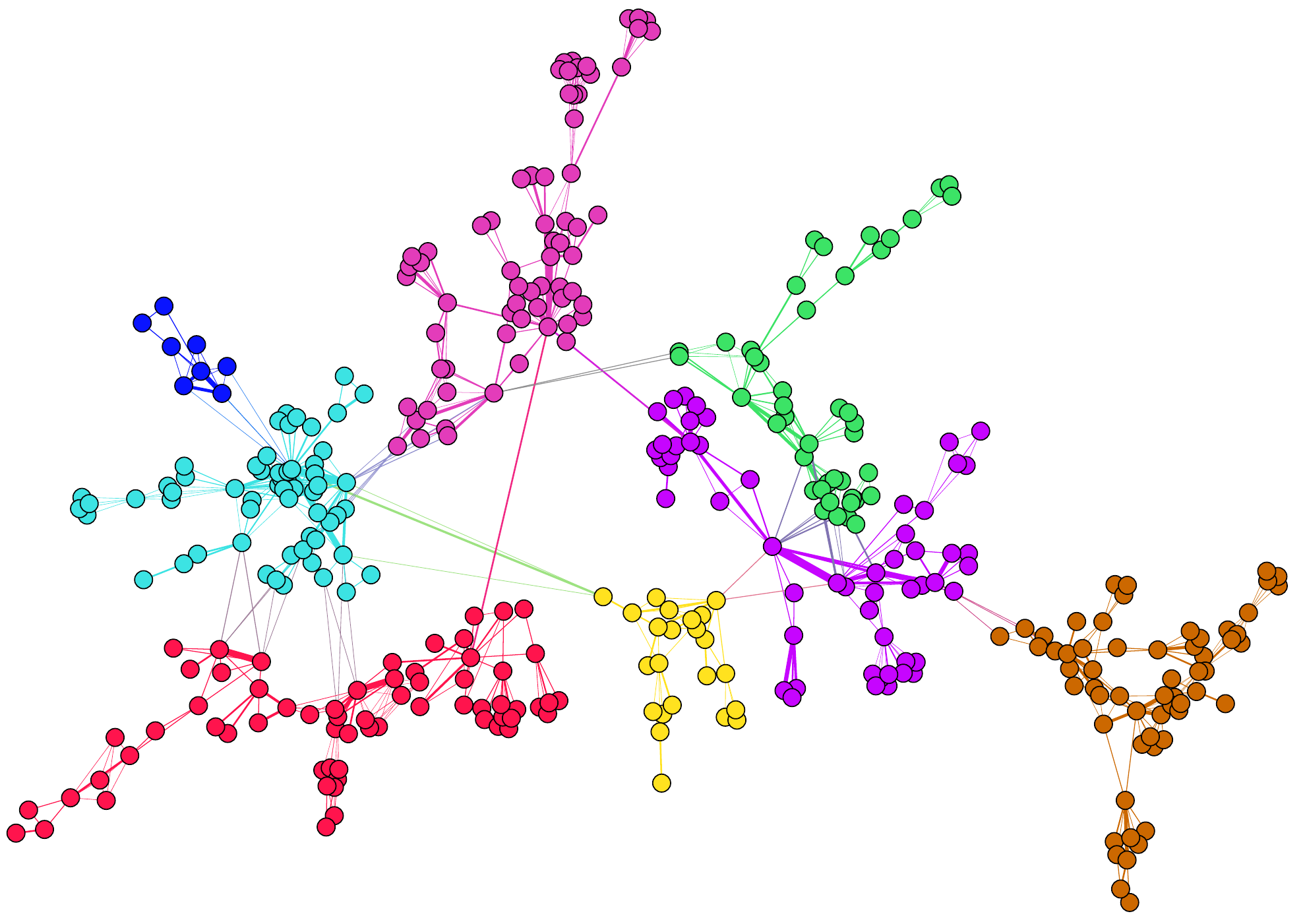}
\end{tabular}
\caption{Graph Clustering of $Dcut$ on the coauthorship network of scientists (K=8). }
\label{fig:netscience}
\end{figure}

\begin{figure}[!tb]
\centering
\begin{tabular}{cccc}
\includegraphics[height=76mm]{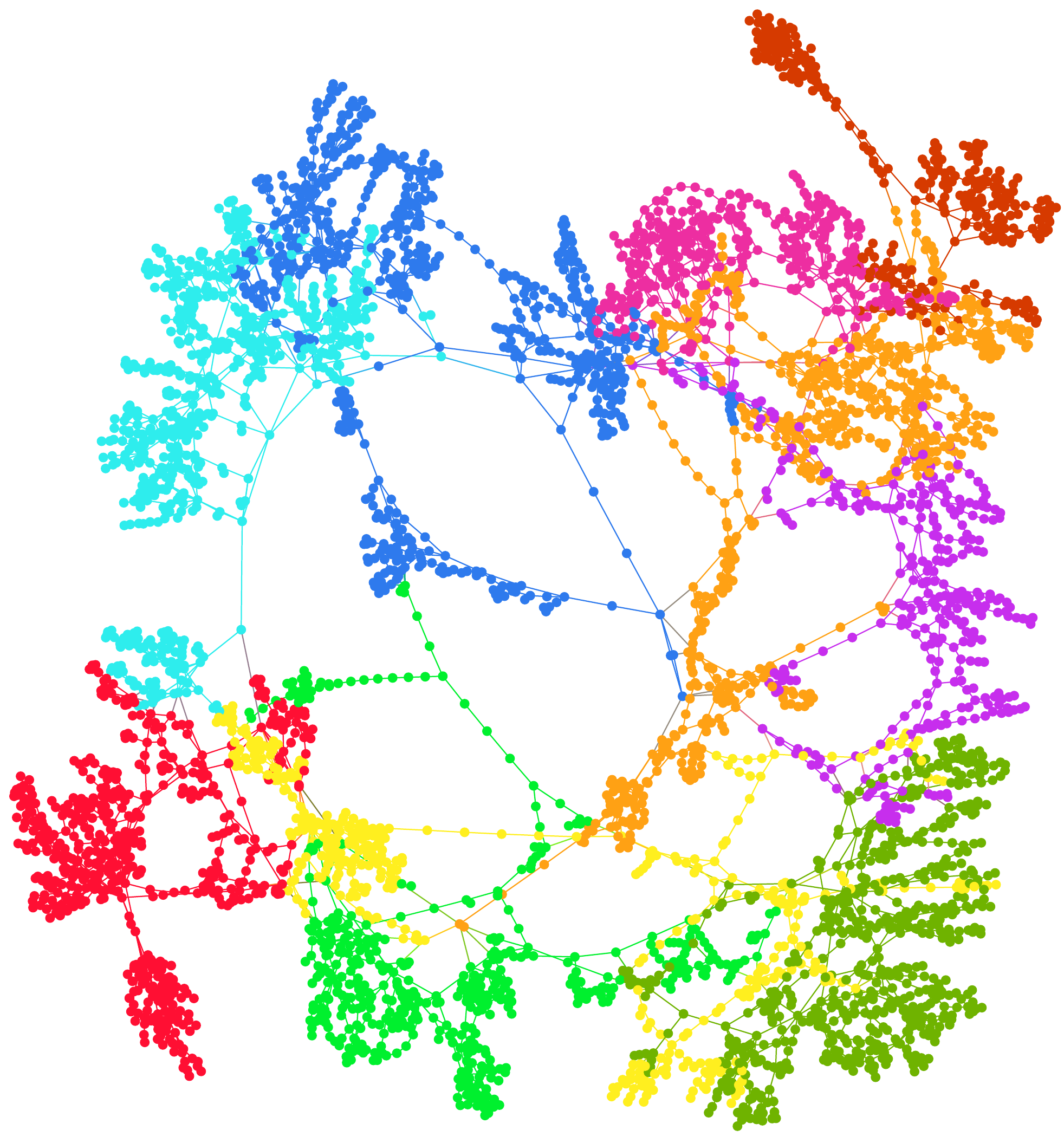}
\end{tabular}
\caption{Graph Clustering of $Dcut$ on the power grid network (K=10). }
\label{fig:power}
\end{figure}

\tabcolsep=4pt
\begin{table}[!t]
\center \caption{Evaluation of different graph clustering algorithms with clustering coefficient on real-world data sets.}
\begin{tabular}{|c|c|c|c|c|c|c|c|c|c|c|c|c|c|c|c|c|c|c|c|c|c|c|c|}
\hline
Data & Dcut & Ncut & Modularity& Metis&MCL\\ \hline
Coauthorship network &0.1408&	0.1349 &	0.1269 &	0.0984 &	0.1247 \\ \hline
Power Grid  & 0.0309& 	0.0255	& 0.0086 &	0.0237	& 0.012\\ \hline
\end{tabular}
\label{tab:CC}
\vspace{-2mm}
\end{table}

\vspace{2mm}
\hspace{-4mm}\textbf{2. Networks without class information}
\vspace{2mm}

 \textbf{Coauthorships in network science:} The graph is a coauthorship network of 1589 scientists working on network theory and experiment. As the vertices of the network are not all connected, only the largest component of this network is used for graph clustering in this study. The graph clusters detected by $Dcut$ (K = 8) are illustrated in Fig. \ref{fig:netscience}.  In the plot,  the obtained clusters present a high degree of scientific community structures. For comparison, the adapted clustering coefficient is applied to measure the quality of graph clusters discovered by different clustering algorithms (Table \ref{tab:CC}).

 \textbf{Power Grid:} This network consists of 4941 vertices and 6594 edges, which represents the power grid of the Western States of the United States, compiled by Duncan Watts and Steven Strogatz. With K = 10, the clustering result of $Dcut$ is depicted in Fig. \ref{fig:power}. We can observe that the power stations in each cluster show strong connections although the graph is very sparse, which results in the highest clustering coefficient of 0.031 compared to other approaches (Table \ref{tab:CC}).

Generally, $Dcut$ allows identifying a good graph clustering, and outperforms the compared algorithms on these real-world data sets, as indicated by external measures or the clustering coefficient.

\begin{figure}[!tb]
\centering
\begin{tabular}{cccc}
\includegraphics[height=45mm]{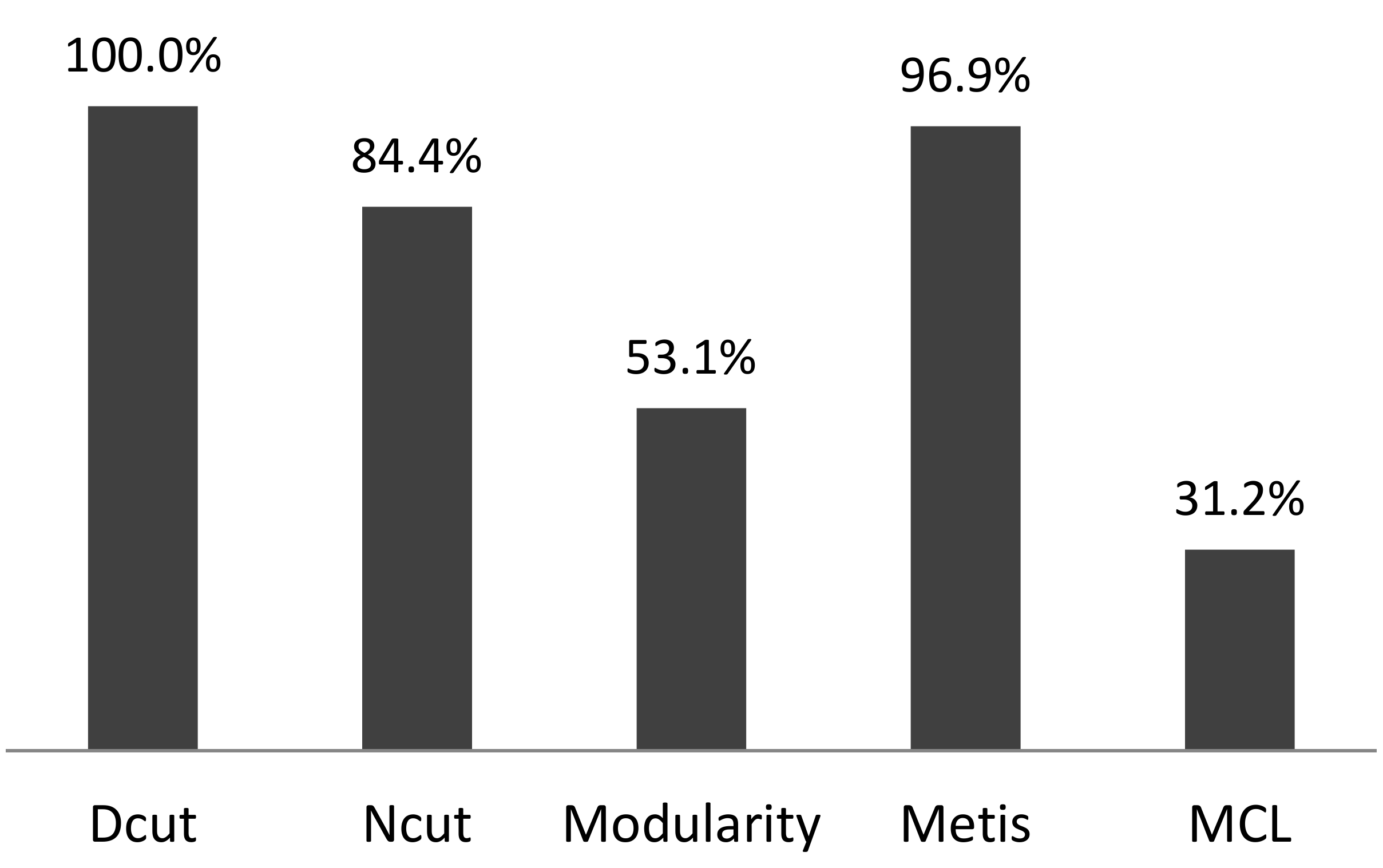}
\end{tabular}
\caption{Performance on protein data set. The clustering results of each algorithm are shown as a bar graph, which measures the percentage of GO enriched clusters with a significance level of $< 0.01$.}
\label{fig:casestudy}
\end{figure}

\subsection{Case Study}
\label{sec:case}
In this section, we further evaluate the performance of $Dcut$ on a case study with a protein-protein interaction (PPI) network. 
Here, we use the PPI network in budding yeast, which contains 2361 proteins and 7182 interactions (http://vlado.fmf.uni-lj.si/pub/networks/data/). We analyze this interaction network with $Dcut$, and also compare its performance to the other four approaches. In the context of biology, we evaluate the biological significance of obtained clusters with the help of the Gene Ontology (GO) database \cite{ontology}, which provides the ontology of defined terms representing functional annotations of proteins. Researchers can calculate P-value of each non-singleton cluster to demonstrate the statistical enrichment of GO molecular functions based on the hyper-geometric distribution \cite{pvalue}.

$Modularity$ detects 32 clusters on this network. For comparison, we set $K = 32$ for $Dcut$, $Ncut$ and $Metis$. It is interesting to observe that all graph clusters detected by $Dcut$ are enriched for the molecular functions (Fig. \ref{fig:casestudy}). For $Metis$ and $Ncut$,  one and five out of 32 clusters are not biologically significant for molecular functions, respectively. $Modularity$ only finds 17 clusters which are enriched for the molecular functions. $MCL$ generats 497 clusters, of which approximately 31\% (159 out of 497) are biologically meaningful clusters.

\subsection{Runtime}
\begin{figure}[!tb]
\centering
\begin{tabular}{cccc}
\includegraphics[height=55mm]{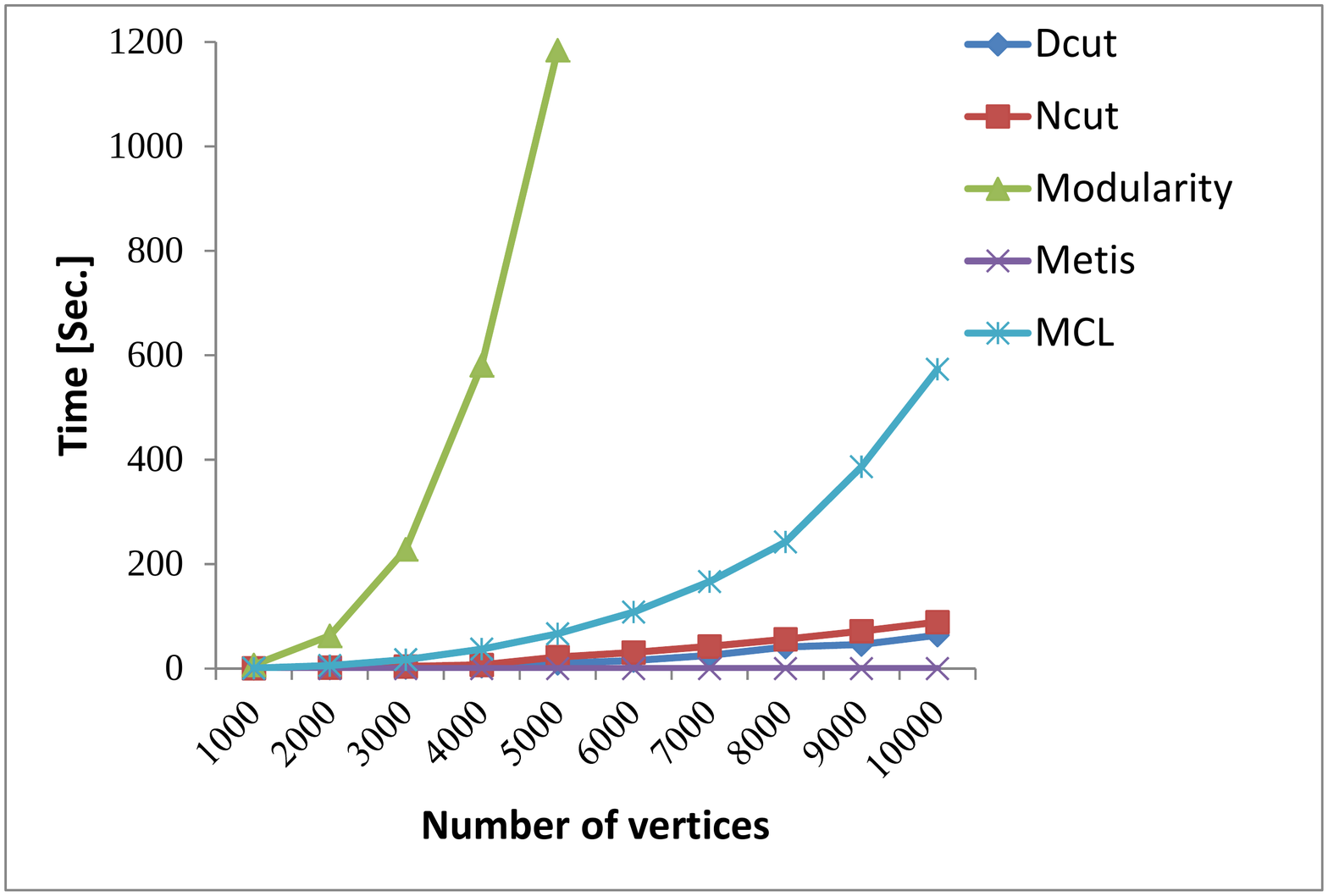}
\end{tabular}
\caption{The runtime of the different graph clustering algorithms. }
\label{fig:time}
\end{figure}

For runtime comparisons, we generated several synthetic data sets,  where the number of clusters $k$ varied from 10 to 50, and each cluster contains 100 nodes. Approximately 30\% of the intra-cluster edges were generated, and 1\% inter-cluster edges were linked. To obtain more accurate runtime results, each method was run 10 times and the times were averaged. In Fig. \ref{fig:time}, we can observe that $Dcut$ is faster than $Modularity$, while the time complexity of $Dcut$ is only super-linear.  $Dcut$ is also better than $MCL$ (approximately 8 times), and comparable to $Ncut$. Moreover, $Dcut$ is slightly slower than $Metis$ with VLSI parallel implementation.

\section{Conclusions}
\label{sec:conclusion}
In this paper, we introduce $Dcut$, a novel graph clustering algorithm. From a density point of view, the proposed \emph{density-cut} criterion offers a more natural and precise measure to quantify the ``goodness" of a graph clustering. Since the constructed density-connected tree of $Dcut$ provides a density connectivity map of vertices in a graph, it supports an efficient way to bisect a graph directly. Our extensive experiments demonstrate that $Dcut$ has many desirable properties and outperforms several state-of-art graph clustering methods.

\end{document}